\documentclass[12pt]{article}
\def\int {\intop \limits}

\def\fnote#1{\footnote}
\begin{document}

\newcommand{\dst}[1]{\displaystyle{#1}}
\newcommand{\barl}{\begin{array}{rl}}
\newcommand{\ball}{\begin{array}{ll}}
\newcommand{\ear}{\end{array}}
\newcommand{\barc}{\begin{array}{c}}
\newcommand{\e}{\mbox{${\bf e}$}}
\newcommand{\J}{\mbox{${\bf J}$}}
\newcommand{\be}{\begin{equation}}
\newcommand{\ee}{\end{equation}}
\newcommand{\aq}[1]{\label{#1}}
\renewcommand \theequation{\thesection.\arabic{equation}}

\title{Variation of radiation length due to LPM effect}
\author{V. N. Baier
and V. M. Katkov\\
Budker Institute of Nuclear Physics\\ 630090 Novosibirsk,
Russia}

\maketitle

\begin{abstract}

The photon emission intensity spectrum is calculated taking into
account influence of multiple scattering (the LPM effect) under
conditions of recent CERN SPS experiment. It is shown that the
theory quite satisfactory describes data.
The integral characteristics: the radiation length and total
probability of photon emission are analyzed under influence of
the LPM effect, and the asymptotic expansions of these characteristics
are derived.

\end{abstract}

\newpage

\section{Introduction}

When a charged particle is moving in a medium it scatters on atoms.
With probability $\sim \alpha$ this scattering is accompanied by a radiation.
At high energy the radiation process occurs over a rather long distance,
known as the {\em formation length} $l_c$ (see e.g.\cite{BKS}):
\begin{equation}
l_c=\frac{l_0}{1+\gamma^2 \vartheta_c^2},\quad
l_0=\frac{2\varepsilon \varepsilon'}{m^2\omega},
\label{1.1a}
\end{equation}
where $\omega$ is the energy of emitted photon,
$\varepsilon (m)$ is the energy
(the mass) of a particle, $\varepsilon'=\varepsilon-\omega$,
$\vartheta_c$ is the characteristic angle of photon emission,
the system $\hbar=c=1$ is used.

Landau and Pomeranchuk were the first who showed that if the formation
length of bremsstrahlung becomes comparable to the distance over which
the multiple scattering becomes important, the bremsstrahlung will be
suppressed \cite{1}. Migdal \cite{M1} developed the quantitative
theory of this phenomenon.

A new interest to the theory of the LPM effect
is connected with a very successful series of experiments performed
at SLAC  \cite{E1}.
In these experiments the cross section
of the bremsstrahlung of soft photons with energy from 200~keV to
500~MeV from electrons with energy 8~GeV and 25~GeV is measured
with an accuracy of the order of a few percent. Both LPM and
dielectric suppression are observed and investigated.
The logarithmic binning of photon spectrum was used
with 25 bins per decade.These experiments were the
challenge for the theory since in all the mentioned papers calculations
are performed to logarithmic accuracy which is not enough for description
of the new experiment. The contribution of the Coulomb corrections (at least
for heavy elements) is larger than experimental errors and these corrections
should be taken into account.

Recently new study of LPM effect at higher energies of electrons
($\varepsilon=$149, 207 and 287~GeV), where the effect has influence upon much
wider part of spectrum comparing with $\varepsilon=$25~GeV,
was performed in the H2 beam line of the CERN SPS \cite{HU},\cite{HU1}.
The logarithmic binning of photon spectrum was used
with 25 bins per decade as at SLAC.

We developed the new approach to the theory of the
Landau-Pomeranchuk-Migdal (LPM) effect \cite{L1}.
In this paper the cross section of the bremsstrahlung process
in the photon energies region where the influence of the LPM is very strong
was calculated with a term $\propto 1/L$ , where $L$
is characteristic logarithm of the problem,
and with the Coulomb corrections
taken into account. In the photon energy region, where the LPM effect
is "turned off", the obtained cross section
gives the exact Bethe-Maximon cross section (within power accuracy and with
the Coulomb corrections). This important feature was absent in
the previous calculations.
The LPM effect in a thin target were the interference effects and
boundary photon emission is of special interest was analyzed
in \cite{L2}. The probability of multiphoton emission is enhanced
at high energy \cite{L3}. Correspondingly this effect is very important
for comparison of theory prediction and data. 
The influence of LPM effect on integral characteristics of
bremsstrahlung was considered in \cite{L4}. 
The other approaches to the LPM effect theory see e.g.
in \cite{Z}, \cite{BDM}, the recent review is given in \cite{BK}.

In Sec.2 the theory predictions are compared with the recent 
CERN SPS data \cite{HU},\cite{HU1}. It is shown that the theory
quite satisfactory describes data. In Sec.3 the variation of the
radiation length due to multiple scattering is discussed. In the region
where the LPM effect is weak ($\varepsilon \ll \varepsilon_e$) 
the asymptotic expressions for the radiation length and the photon
emission probability are derived taking into account both decomposition
over $\varepsilon /\varepsilon_e$ and over $1/L_1,~L_1$ is the 
characteristic logarithm of the problem. Appendices A and B contains
details of derivation for radiation length and Appendix C for the photon
emission probability. In Appendix D the structure of series 
over $\varepsilon /\varepsilon_e$ and over $1/L_1$ is analyzed
for $\varepsilon \ll \varepsilon_e$.

\section{Influence of the multiple scattering
on the bremsstrahlung spectrum}
\setcounter{equation}{0}

The spectral radiation intensity obtained in \cite{L1}, Eq.(2.39)
(see also Sec.3 in \cite{L4}) is valid for any energy and has the form
\begin{equation}
dI=\omega dW=\frac{\alpha m^2 xdx}{2\pi (1-x)}
{\rm Im}~\left[\Phi(\nu)-\frac{1}{2L_c}F(\nu)
\right],\quad x=\frac{\omega}{\varepsilon},
\label{3.1}
\end{equation}
where
\begin{eqnarray}
&&\displaystyle{\Phi(\nu)=\int_{0}^{\infty} dz e^{-it}\left[r_1
\left(\frac{1}{\sinh z}-\frac{1}{z}\right)-i\nu r_2
\left( \frac{1}{\sinh^2z}- \frac{1}{z^2}\right) \right]}
\nonumber \\
&& = r_1\left(\ln p-\psi\left(p+\frac{1}{2}\right) \right)
+r_2\left(\psi (p) -\ln p+\frac{1}{2p}\right),
\nonumber \\
&&F(\nu)= \int_{0}^{\infty}\frac{dz e^{-it}}{\sinh^2z}
\left[r_1f_1(z)-2ir_2f_2(z) \right],
\nonumber \\
&& t=\frac{z}{\nu},\quad r_1=x^2,\quad r_2=1+(1-x)^2.
\label{3.2}
\end{eqnarray}
where $z=\nu t,~p=i/(2\nu),~\psi(x)$ is the logarithmic derivative of
the gamma function.
The functions $f_1(z)$ and $f_2(z)$ are defined by following
expressions
\begin{eqnarray}
&& f_1(z)=\left(\ln \varrho_c^2+\ln \frac{\nu}{i}
-\ln \sinh z-C\right)g(z) - 2\cosh z G(z),
\nonumber \\
&& f_2(z) = \frac{\nu}{\sinh z}
\left(f_1(z)-\frac{g(z)}{2} \right),\quad
 g(z)=z\cosh z - \sinh z,
\nonumber \\
&& G(z)=\int_{0}^{z}(1-y\coth y)dy
\nonumber \\
&&\displaystyle{=z-\frac{z^2}{2}-\frac{\pi^2}{12}-
z\ln \left(1-e^{-2z} \right)
+\frac{1}{2}{\rm Li}_2 \left(e^{-2z} \right)},
\label{2.15}
\end{eqnarray}
here ${\rm Li}_2 \left(x \right)$ is the Euler dilogarithm.
Use of the last representation of function $G(z)$ simplifies
the numerical calculation. The crucial parameter $\nu$ is
\begin{eqnarray}
&& \nu^2=i\nu_0^2,\quad \nu_0^2=|\nu|^2 \simeq \nu_1^2\left(1+
\frac{\ln \nu_1}{L_1}\vartheta(\nu_1-1) \right),\quad
\nu_1^2=\frac{\varepsilon}{\varepsilon_e}\frac{1-x}{x},
\nonumber \\
&&\varepsilon_e=m\left(8\pi Z^2 \alpha^2 n_a \lambda_c^3 L_1 \right)^{-1},
\quad L_c \simeq L_1 \left(1+
\frac{\ln \nu_1}{L_1}\vartheta(\nu_1-1) \right),\quad
L_1=\ln \frac{a_{s2}^2}{\lambda_c^2},
\nonumber \\
&& \frac{a_{s2}}{\lambda_c}=183Z^{-1/3}{\rm e}^{-f},\quad
f=f(Z\alpha)=(Z\alpha)^2\sum_{k=1}^{\infty}\frac{1}{k(k^2+(Z\alpha)^2)},
\label{3.3}
\end{eqnarray}
where $Z$ is the charge of the nucleus, $n_a$ is the number density
of atoms in the medium, $\lambda_c=1/m=(\hbar/mc)$ is the electron Compton
wavelength. Here $\varepsilon_e$ is a characteristic parameter of medium,
starting from this energy the multiple scattering distorts the whole
spectrum of bremsstrahlung including its hard part: for iridium
$\varepsilon_e$=2.27~TeV, for gold $\varepsilon_e$=2.6~TeV,
for tungsten $\varepsilon_e$=2.73 ~TeV,
for tantalum $\varepsilon_e$=3.18 ~TeV,
for lead $\varepsilon_e$=4.38~TeV.

In the case $\varepsilon \ll \varepsilon_e$ the LPM effect manifests
itself when
\begin{equation}
\nu_1(x_c)=1,\quad x_c=\frac{\varepsilon}{\varepsilon_e+\varepsilon} \simeq \frac{\varepsilon}{\varepsilon_e}.
\label{3.4}
\end{equation}
In this case in the hard part of spectrum ($1 \geq x \gg x_c,~
\nu_1^2 \simeq x_c/x \ll 1$) one has (see Appendices A and B)
\begin{eqnarray}
&& \frac{dI}{dx}=\frac{\varepsilon}{L_{rad}^0}\Bigg\{x^2 +\frac{4(1-x)}{3}
+\frac{2(1-x)}{9L_1} - \frac{\varepsilon^2}{\varepsilon_e^2} \frac{(1-x)^2}{x^2}
\nonumber \\
&& \times \left[\left(\frac{64}{63}-\frac{15272}{2205}\frac{1}{L_1} \right)(1-x)
+ \left(1-\frac{5017}{1800}\frac{1}{L_1} \right)x^2 \right]\Bigg\}, 
\nonumber \\
&& \frac{1}{L_{rad}^0}=\frac{2Z^2\alpha^3n_a L_1}{m^2}
=\frac{\alpha}{4\pi}\frac{m}{\varepsilon_e\lambda_c},\quad \frac{1}{L_{rad}^{BM}}=
\frac{1}{L_{rad}^0}\left(1+\frac{1}{9L_1} \right), 
\label{3.6}
\end{eqnarray}
here $L_{rad}^{BM}$ is the Bethe-Maximon radiation length. 
Note that if neglect here the terms 
$\propto \varepsilon^2/\varepsilon_e^2$ we obtain
the Bethe-Maximon intensity spectrum.

In the recent CERN SPS experiment
the iridium target with thickness $l$=0.128~mm
($ l/L_{rad}=4.36 \pm 0.10 \%,~L_{rad}$ is the radiation length)
\cite{HU} and tantalum target \cite{HU1} with thickness
$ l/L_{rad}=4.45 \pm 0.10 \%$ were used.
Photons with energy 2~GeV $< \omega < \varepsilon$  were detected in a
lead glass calorimeter.

Obtained experimental data should be recalculated:
\begin{equation}
\left(\frac{d\varepsilon}{d\omega}\right)_{exp}=
\frac{l}{L_{rad}}\frac{1}{k} \left(
\frac{dN}{d\ln\omega}\right)_{exp}.
\label{D2}
\end{equation}
Because photon energies were histogrammed logarithmically,
using 25 bins per decade of energy, one has for the coefficient $k$
\begin{eqnarray}
&&k_{h}=\frac{\omega_{max}-\omega_{min}}{\omega_{min}}
=\displaystyle{e^s-1}=0.096,\quad
k_{m}=2\frac{\omega_{max}-\omega_{min}}{\omega_{max}+\omega_{min}}
=2\frac{e^s-1}{e^s+1}=0.092,
\nonumber \\
&&k_{l}=\frac{\omega_{max}-\omega_{min}}{\omega_{max}}=1-e^{-s}=0.88,\quad
s=\ln\frac{\omega_{max}}{\omega_{min}}=\frac{\ln 10}{25}=0.092,
\label{D3}
\end{eqnarray}
depending on the normalization point within bin.
Here we will use $k_{m}$.

Since only high energy photons ($\omega \geq 2~$GeV) were measured,
no boundary effects were observed. So, only the pure
LPM effect was studied. For iridium $\varepsilon_e$=2.27~TeV and
from one has Eq.(\ref{3.4}) that the characteristic photon energy
$\omega_c(\varepsilon)$ for which the LPM effect is well
manifests itself is $\omega_c$(287~GeV)=32~GeV and
$\omega_c$(207~GeV)=19~GeV, while for tantalum
$\varepsilon_e$=3.18 ~TeV and  $\omega_c$(287~GeV)=26~GeV.
The results of calculations for iridium (the initial electron
energy $\varepsilon$=287~GeV and $\varepsilon$=207~GeV) and for
tantalum (the initial electron energy $\varepsilon$=287~GeV)
are shown in Figs.1-3. The curve 1 is the Bethe-Maximon intensity
spectrum (see e.g.Eq.(\ref{3.6})). The curves  2, 3 are
calculated using Eqs. (\ref{3.1}) and (\ref{3.2}). The curve 2
presents the main term (the function $\Phi(\nu)$), the curve 3
presents the correction term (the function $F(\nu)$).  It should be
noted that the prediction of our theory (sum of
previous terms, curve 4) in the hard end of spectrum coincide
with the Bethe-Maximon curve within the accuracy
better than $10^{-3}$. It is seen that given values
$\omega_c(\varepsilon)$ show the scale where LPM effect becomes
essential. For used thickness
of target the multi-photon effects are very essential. The reduction
factor $f$ in this case it is convenient to calculate using
the following general expression (Eq.(3.4) of \cite{L3}):
\begin{equation}
\frac{d\varepsilon}{d\omega}=\omega \frac{dw}{d\omega} f,\quad
f= \exp \left[-\int_{\omega}^{\infty}\frac{dw}{d\omega_1}d\omega_1 \right]
\left(1+O\left(\omega \frac{dw}{d\omega} \right) \right),
\label{y3.5}
\end{equation}
where the main term (with $\Phi(\nu)$) of spectrum Eq.(\ref{3.1})
is substituted. The results obtained are in a good agreement
with Eqs.(2.11) and (2.26) of \cite{L3}. The reduction factor $f$
in iridium for three used energies is presented in Fig.4.
It  follows from Fig.4 that if the electron energy decreases
the reduction factor $f$ diminishes as well.
The final prediction, with the reduction factor taken
into account for used target thicknesses,
present the curves $T$. The data are recalculated according with
Eqs.(\ref{D2}),(\ref{D3}) using the coefficient $k_m$.
In Figs.1-3 one can see that for energy 287 GeV there is quite satisfactory
agreement of theory with data for both iridium and tantalum,
for energy 207 GeV in iridium the
agreement is somewhat less satisfactory.

\section{The LPM radiation length and the average number of
emitted photons at $\varepsilon \ll \varepsilon_e$}
\setcounter{equation}{0}

The local radiation length $L_{rad}$ is defined by equation
\begin{equation}
\frac{d\varepsilon}{dt}= -I(\varepsilon) = -\frac{\varepsilon}{L_{rad}}.
\label{4.1}
\end{equation}
In the absence of LPM effect the value $L_{rad}$ doesn't dependent 
on energy and
defined by Eq.(\ref{3.6}). In this case, after averaging Eq.(\ref{4.1}) with
the electron distribution function over energy $f(\varepsilon, t)$ one obtains
the closed equation for the mean energy loss which has the solution 
$\left<   \varepsilon(t) \right>  =\varepsilon_0 \exp(-t/L_{rad})$.

For the first photon emission according with Eq.(\ref{4.1}) 
$L_{rad}=\varepsilon/I(\varepsilon)$. The dependence on the energy of the function
$(I(\varepsilon)/\varepsilon) L_{rad}^{BM}=L_{rad}^{BM}/L_{rad}$ for iridium 
(curve 1) and lead (curve 2) is given in Fig.5. The $I(\varepsilon)$ is the 
integrated over $x$ Eq.(\ref{3.1}). The relative value of correction (the
term with $F(\nu)$) to the main term (with $\Phi(\nu)$) depends on energy. 
It attains the maximum $\sim 5\%$ near $\varepsilon \sim \varepsilon_e$
for both media. So, the value of $L_{rad}$ for the energy $\varepsilon=3~$TeV
increases 1.33 times in Ir and in 1.16 times in Pb, while  
for the energy $\varepsilon=10~$TeV the corresponding figures are 1.58 and
1.38.

In the region where the LPM effect is weak (see Appendices A and B)
\begin{eqnarray}
&& \frac{1}{L_{rad}}=\frac{1}{L_{rad}^0}\Bigg\{1+\frac{1}{9L_1}
-\frac{4\pi}{15}\frac{\varepsilon}{\varepsilon_e} +\frac{64}{21}
\frac{\varepsilon^2}{\varepsilon_e^2}\left(\ln\frac{\varepsilon_e}{\varepsilon}
-2.040\right) 
\nonumber \\
&& +\frac{1}{L_1}\left[\frac{182\pi}{225}\frac{\varepsilon}{\varepsilon_e}-
\frac{15272}{735}\frac{\varepsilon^2}{\varepsilon_e^2}
\left(\ln\frac{\varepsilon_e}{\varepsilon} - 2.577\right)  \right]. 
\label{4.2}
\end{eqnarray}
The first term of expansion over $\varepsilon/\varepsilon_e$ (not over $1/L_1$)
was found in \cite{L4}. 

The accuracy of Eq.(\ref{4.2}) is defined by a several
essentially different factors. The first is the relative value of discarded 
terms $O(\varepsilon^3/\varepsilon_e^3)$ in Eqs.(\ref{a.13}), \ref{b.15}).
The second is the substitution of $L_c$ by $L_1~(\varrho_c=1)$ in 
Eqs.(\ref{3.1}) - (\ref{3.3}) for the whole spectrum, the relative accuracy 
of this substitution is $O(\varepsilon/(\varepsilon_e L_1^2))$. 
The third is due to fact that in the initial
formula Eq.(\ref{3.1}) the terms $\propto 1/L_c^2$ were rejected. However, 
as is shown in in Appendix D, in the region $\varepsilon \ll \varepsilon_e$
the corrections $\propto 1/L_1^2$ are beginning with the terms contained 
$\varepsilon/\varepsilon_e$. So, in this region, the relative accuracy in the
formula Eq.(\ref{3.1}) is
also $O(\varepsilon/(\varepsilon_e L_1^2))$.

One can see from Eq.(\ref{4.2}), that in a decomposition over 
$\varepsilon/\varepsilon_e$ one have to take into account the terms
$\propto 1/L_1$ because of relatively large numerical values of the
coefficients of decomposition. Moreover, for heavy elements 
$L_1 \simeq (6 \div 7)$ taking into account of term $\propto 1/L_1$ results
in substantial change of numerical value of coefficients at the given
degree of $\varepsilon/\varepsilon_e$. Particularly this is important for
term containing $\ln (\varepsilon_e/\varepsilon)$. As a result, there is
essential compensation inside the coefficients of decomposition. This
permit to use the decomposition Eq.(\ref{4.2}) at relatively large energies.
At $\varepsilon/\varepsilon_e \leq 1/10~$ an error in the correction terms
doesn't exceed 10\%.

As illustration we present the radiation length in iridium ($L_1=6.9225,
~\varepsilon_e=2.27~$TeV,~$L_{rad}^{BM}$(Ir)=2.91~mm):
\begin{equation}
\frac{1}{L_{rad}({\rm Ir})}=\frac{1}{L_{rad}^{BM}({\rm Ir})}\left(1-0.464 
\frac{\varepsilon}{\varepsilon_e} +0.045\frac{\varepsilon^2}
{\varepsilon_e^2}\ln \frac{\varepsilon_e}{\varepsilon}
+1.492\frac{\varepsilon^2}{\varepsilon_e^2}\right). 
\label{4.3}
\end{equation} 
It should be noted that after substitution of Eq.(\ref{4.3}) into Eq.(\ref{4.1}) 
and averaging with the distribution function $f(\varepsilon, t)$ the higher 
moments of $\varepsilon$ appear. Because of this, equation (\ref{4.1}) 
ceases to be closed. We will discuss this item elsewhere.

The mean value of photons emitted by electron is also of evident interest. 
Particularly, at $\omega \ll \omega_c \simeq \varepsilon^2/\varepsilon_e$ 
this value defines mainly the exponent in Eq.(\ref{y3.5}). 
At energy where the LPM effect is rather weak 
($\varepsilon \ll \varepsilon_e$) the radiation probability per unit
time integrated over whole photon spectrum is (see Appendix C)
\begin{eqnarray}
&& W(\varepsilon)=\frac{dw}{dt}=\frac{\alpha m}{3\pi \varepsilon_e \lambda_c}
\Bigg[\ln \frac{\varepsilon_e}{\varepsilon}+1.959 +\frac{2\pi}{5}
\frac{\varepsilon}{\varepsilon_e}
\nonumber \\
&& +\frac{1}{L_1}\left(\frac{1}{6}\ln \frac{\varepsilon_e}{\varepsilon}
+3.574 - \frac{91\pi}{75}\frac{\varepsilon}{\varepsilon_e}\right) \Bigg]. 
\label{4.4}
\end{eqnarray}
It is seen that allowing for terms $\propto 1/L_1$ changes
essentially non-logarithmic terms in Eq.(\ref{4.4}).

\vspace{0.5cm}

{\bf Acknowledgments}

The authors are indebted to the Russian Foundation for Basic
Research supported in part this research by Grant 
03-02-16154.

We would like to thank U.Uggerhoj for information about experiment
and data.

\newpage

\setcounter{equation}{0}
\Alph{equation}
\appendix

\section{Appendix}

We consider here the asymptotic behavior of the total
intensity of radiation $I_0$ in the region
$\varepsilon \ll \varepsilon_e$ where the LPM effect
is weak and one can put in Eqs.(\ref{3.1})-(\ref{3.3})
$\nu^2=i\nu_1^2, \varrho_c=1, L_c=L_1$. This substitution will
manifest itself only starting from terms of the order
$\displaystyle{\frac{1}{L_1^2}\frac{\varepsilon}{\varepsilon_e}I_0}$.
Then in expressions for $\Phi(\nu)$ and $F(\nu)$ in Eq.(\ref{3.2})
we integrate by parts the terms $\propto r_1$ and make the substitution
of variable $t \rightarrow -it$. As a result we obtain
\begin{equation}
I=\frac{\varepsilon}{L_{rad}^0}\left(J_m +\frac{1}{L_1}J_v \right),
\label{a.1}
\end{equation}
where
\begin{eqnarray}
&& J_m=2{\rm Re}\int_{0}^{1}dx\int_{0}^{\infty}dt e^{-t}
\left[x^2\varphi_1(z)-2(1-x)\varphi_2(z)\right];
\nonumber \\
&& \varphi_1(z)=\frac{1}{\cosh z+1},\quad
\varphi_2(z)=\frac{1}{\sinh^2 z}-\frac{1}{z^2},
\label{a.2}
\end{eqnarray}
and
\begin{eqnarray}
&& J_v=-{\rm Re}\int_{0}^{1}dx\int_{0}^{\infty}dt e^{-t}
\left[x^2l_1(z, t)+4(1-x)l_2(z, t)\right];
\nonumber \\
&& l_{1,2}=a_{1,2}(z)(\ln t +C) + b_{1,2}(z),~
a_1(z)=\varphi_1(z)-\frac{z}{\sinh z},~
a_2(z)=\frac{z \coth z -1}{\sinh^2 z},
\nonumber \\
&& b_1(z)=a_1(z) \ln\frac{\sinh z}{z}+ \frac{1}{\cosh z+1}
\left(1-z\coth z +\frac{2}{\sinh z} G(z) \right),
\nonumber \\
&& b_2(z)=a_2(z) \ln\frac{\sinh z}{z}+ \frac{1}{2\sinh^2 z}
\left(4 G(z)\coth z+z\coth z -1\right),
\label{a.3}
\end{eqnarray}
here the function $G(z)$ is defined in Eq.(\ref{2.15}) and
\begin{equation}
\beta^2=i\frac{\varepsilon_e}{\varepsilon},\quad
z=\frac{t}{\beta y},\quad y=\sqrt{\frac{x}{1-x}}.
\label{a.4}
\end{equation}
In Eq.(\ref{a.2}) one can expand the function $\varphi_1(z)$
up to terms $\propto z^4$. We find
\begin{equation}
J_{m1} \simeq 2\int_{0}^{1}dx x^2\int_{0}^{\infty}dt e^{-t}
\left(\varphi_1(0)+ \frac{\varphi_1^{(4)}(0)}{\beta^4 y^4}
\frac{t^4}{4!}\right)=\frac{2}{3}\left[\varphi_1(0)-
\frac{\varepsilon^2}{\varepsilon_e^2} \varphi_1^{(4)}(0)\right].
\label{a.5}
\end{equation}
In the integral with function $\varphi_2(z)$ in Eq.(\ref{a.2})
we perform twice integration by parts over $t$. We find
\begin{eqnarray}
&& J_{m2}=-4{\rm Re}\int_{0}^{1}dx(1-x)\left[\varphi_2(0)+
\frac{1}{\beta^2y^2}\int_{0}^{\infty}dt e^{-t}\varphi_2^{(2)}(z) \right]
=-2\varphi_2(0) -4\frac{\varepsilon}{\varepsilon_e}S,
\nonumber \\
&& S={\rm Im}\int_{0}^{1}\frac{(1-x)^2 dx}{x}\int_{0}^{\infty}dt
e^{-t}\varphi_2^{(2)}(z).
\label{a.6}
\end{eqnarray}
To calculate $S$ we choose the value $y_0$ in such way that $y_0 \ll 1$
and $|\beta|y_0 \gg 1$ and divide the integral over $x$ into two.
The first one is
\begin{eqnarray}
\hspace{-15mm}&& S_1={\rm Im}\int_{x_0}^{1}\frac{(1-x)^2 dx}{x}
\int_{0}^{\infty}dt e^{-t}\varphi_2^{(2)}(z) \simeq
{\rm Im}\int_{x_0}^{1}\frac{(1-x)^2 dx}{x}\frac{1}{\beta^2y^2}
\varphi_2^{(4)}(0)
\nonumber \\
\hspace{-15mm}&& \simeq -\frac{1}{|\beta|^2}\left(\frac{1}{x_0}-1+3\ln x_0+
\frac{5}{2} \right) \varphi_2^{(4)}(0) \simeq
-\frac{1}{|\beta|^2}\left(\frac{1}{y_0^2}+6\ln y_0+
\frac{5}{2} \right) \varphi_2^{(4)}(0)
\label{a.7}
\end{eqnarray} 
In the second integral we pass to the variable $y$:
\begin{equation}
S_2=2{\rm Im}\int_{0}^{y_0}\frac{dy}{y(1+y^2)^3}
\int_{0}^{\infty}dt e^{-t}\varphi_2^{(2)}(z) \simeq
2{\rm Im}\int_{0}^{y_0}(1-3y^2)\frac{dy}{y}
\int_{0}^{\infty}dt e^{-t}\varphi_2^{(2)}(z)
\label{a.8}
\end{equation}
Let us consider the integral
\begin{eqnarray}
&& S_{21}=2{\rm Im}\int_{0}^{y_0}\frac{dy}{y}
\int_{0}^{\infty}dt e^{-t}\varphi_2^{(2)}\left(\frac{t}{\beta y}\right)
=\frac{1}{i} \int_{\beta^{\ast}y_0}^{\beta y_0}\frac{dy}{y}
\int_{0}^{\infty}dt e^{-t}\varphi_2^{(2)}\left(\frac{t}{y}\right)
\nonumber \\
&& \simeq \frac{1}{i} \int_{\beta^{\ast}y_0}^{\beta y_0}\frac{dy}{y}
\left(\varphi_2^{(2)}(0)+\frac{1}{y^2}\varphi_2^{(4)}(0) \right)
=\frac{1}{i}\left[\ln\frac{\beta}{\beta^{\ast}}\varphi_2^{(2)}(0)
-\frac{1}{2}\left(\frac{1}{\beta^2y_0^2} -
\frac{1}{\beta^{{\ast}2}y_0^2}\right)
\varphi_2^{(4)}(0) \right]
\nonumber \\
&&=\frac{\pi}{2}\varphi_2^{(2)}(0)+ \frac{1}{|\beta|^2y_0^2}\varphi_2^{(4)}(0).
\label{a.9}
\end{eqnarray}
The remaining integral in Eq.(\ref{a.8}) is
\begin{eqnarray}
&& S_{22}= -6 {\rm Im}\int_{0}^{y_0}ydy
\int_{0}^{\infty}dt e^{-t}\varphi_2^{(2)}(z)
=\frac{6}{|\beta|^2}{\rm Re}\int_{0}^{\beta y_0}\frac{dy}{y}
\int_{0}^{\infty}dt e^{-t}\varphi_2^{(4)}\left( \frac{t}{y}\right)
\nonumber \\
&&=\frac{6}{|\beta|^2}{\rm Re}\left[\ln(\beta y_0)
\int_{0}^{\infty}dt e^{-t}\varphi_2^{(4)}\left( \frac{t}{\beta y_0}\right)
+ \int_{0}^{\beta y_0}\ln ydy \int_{0}^{\infty}\frac{tdt}{y^2} e^{-t}
\varphi_2^{(5)}\left( \frac{t}{y}\right) \right]
\nonumber \\
&& \simeq \frac{6}{|\beta|^2}{\rm Re}\left[\ln(\beta y_0)
\varphi_2^{(4)}(0) + \int_{0}^{\infty}\ln ydy\int_{0}^{\infty}
ze^{-zy}\varphi_2^{(5)}(z)dz\right]
\nonumber \\
&& =\frac{6}{|\beta|^2}{\rm Re}\left[\ln(\beta y_0)
\varphi_2^{(4)}(0) + \int_{0}^{\infty}dy\int_{0}^{\infty}
(\ln y - \ln z)e^{-y}\varphi_2^{(5)}(z)dz\right]
\nonumber \\
&&= \frac{6}{|\beta|^2}{\rm Re}\left[(\ln(\beta y_0)+C)
\varphi_2^{(4)}(0) - \int_{0}^{\infty}
 \ln z\varphi_2^{(5)}(z)dz\right]
\label{a.10}
\end{eqnarray}
Summing $S_1$, $S_{21}$ and $S_{22}$ we obtain for $S$
\begin{equation}
S=\frac{\pi}{2}\varphi_2^{(2)}(0)+3\frac{\varepsilon}{\varepsilon_e}
\left[\left( \ln \frac{\varepsilon_e}{\varepsilon}+2C-\frac{5}{6}\right)
\varphi_2^{(4)}(0) -2 \int_{0}^{\infty}
 \ln z\varphi_2^{(5)}(z)dz\right].
 \label{a.11}
\end{equation}
Substituting  Eq.(\ref{a.11}) into $J_{m2}$, adding with $J_{m1}$
Eq.(\ref{a.5}) and taking into account that
\begin{eqnarray}
&&\varphi_1(0)=\varphi_1^{(4)}(0)=\frac{1}{2},~
\varphi_2(0)=-\frac{1}{3},~\varphi_2^{(2)}(0)=\frac{2}{15},~
\varphi_2^{(4)}(0)=-\frac{16}{63},
\nonumber \\
&& \int_{0}^{\infty}\ln z\varphi_2^{(5)}(z)dz=-0.286,
\label{a.12}
\end{eqnarray}
we obtain
\begin{eqnarray}
\hspace{-10mm}&& J_m=1-\frac{4\pi}{15}\frac{\varepsilon}{\varepsilon_e} +
\frac{64}{21}\frac{\varepsilon^2}{\varepsilon_e^2}\left(
\ln \frac{\varepsilon_e}{\varepsilon}+2C-\frac{5}{6}- \frac{7}{64}
+\frac{63}{8}\int_{0}^{\infty}\ln z\varphi_2^{(5)}(z)dz\right)+
O\left(\frac{\varepsilon^3}{\varepsilon_e^3} \right)
\nonumber \\
\hspace{-10mm}&&= 1-\frac{4\pi}{15}\frac{\varepsilon}{\varepsilon_e}
+\frac{64}{21}\frac{\varepsilon^2}{\varepsilon_e^2}
\left(\ln \frac{\varepsilon_e}{\varepsilon}-2.040 \right)+
O\left(\frac{\varepsilon^3}{\varepsilon_e^3} \right).
\label{a.13}
\end{eqnarray}
Here the first two terms agree with Eq.(3.9) in \cite{L4}.	

\setcounter{equation}{0}

\section{Appendix}

Here we calculate the function $J_v$ in Eq.(\ref{a.1}) defined in
Eq.(\ref{a.3}). In the last expression one can expand the function $l_1(z,t)$
up to terms $\propto z^4$. We find
\begin{eqnarray}
\hspace{-15mm}&& J_{v1} \simeq \left(\frac{\varepsilon}{\varepsilon_e}\right)^2
\int_{0}^{1} (1-x)^2dx \int_{0}^{\infty}dt e^{-t}
\frac{t^4}{4!}\left(a_1^{(4)}(0)(\ln t+C)+ b_1^{(4)}(0)\right)
\nonumber \\
\hspace{-15mm}&& =\frac{1}{3}\left(\frac{\varepsilon}{\varepsilon_e}\right)^2
\left[(\psi(5)-\psi(1))a_1^{(4)}(0)+ b_1^{(4)}(0)\right] =
\frac{1}{3}\left(\frac{\varepsilon}{\varepsilon_e}\right)^2
\left[\frac{25}{12}a_1^{(4)}(0)+ b_1^{(4)}(0)\right],
\label{b.1}
\end{eqnarray}
where the function $\psi(x)$ is defined in Eq.(\ref{3.2}). The integral
with the function $l_2(z, t)$ is
\begin{equation}
J_{v2}=-4{\rm Re}\int_{0}^{1}(1-x)dx\int_{0}^{\infty}dt e^{-t}
\left[a_{2}(z)(\ln t +C) + b_{2}(z)\right].
\label{b.2}
\end{equation}
The term with $b_2(z)$ can be calculated as the term $J_{m2}$ in the Appendix A.
Substituting $\varphi_2(z) \rightarrow b_2(z)$ we have
\begin{eqnarray}
&&J_{v2}^{(b)}=-2b_2(0)-2\pi\frac{\varepsilon}{\varepsilon_e}b_2^{(2)}(0)
\nonumber \\
&&-12 \left(\frac{\varepsilon}{\varepsilon_e}\right)^2\left[
\left(\ln \frac{\varepsilon_e}{\varepsilon} +2C-\frac{5}{6}\right)
b_2^{(4)}(0) -2\int_{0}^{\infty}\ln z b_2^{(5)}(z)dz \right].
\label{b.3}
\end{eqnarray}
Now we turn over to the term with $a_2(z)$ in Eq.(\ref{b.2})
\begin{eqnarray}
&& T={\rm Re}\int_{0}^{1}(1-x)dx\int_{0}^{\infty}dt e^{-t}
a_{2}(z)(\ln t +C)
\nonumber \\
&&={\rm Re}\int_{0}^{1}(1-x)dx\int_{0}^{\infty}dt e^{-t}
\left( a_{2}(z)-a_{2}(0)\right) (\ln t +C).
\label{b.4}
\end{eqnarray}
As in previous Appendix we divide the integral over $x$ into two:
$0 \leq x \leq x_0$ and $x_0 \leq x \leq 1$. We have (see Eqs.(\ref{b.1})
and (\ref{a.7}))
\begin{eqnarray}
\hspace{-15mm}&& T_1={\rm Re}\int_{x_0}^{1}(1-x)dx\int_{0}^{\infty}dt e^{-t}
\left( a_{2}(z)-a_{2}(0)\right) (\ln t +C)
\nonumber \\
\hspace{-15mm}&& =-\frac{1}{|\beta|^4}\int_{x_0}^{1}\frac{(1-x)^3}{x^2}dx
\frac{25}{12}a_1^{(4)}(0) \simeq -\frac{25}{12|\beta|^4}
\left(\frac{1}{y_0^2}+6\ln y_0 +\frac{5}{2} \right) a_2^{(4)}(0).
\label{b.5}
\end{eqnarray}
In the second integral we perform twice integration by parts over $t$
and pass to the variable $y$. We find
\begin{eqnarray}
&& T_2 = \frac{2}{|\beta|^2} {\rm Im}\int_{0}^{y_0}\frac{dy}{y(1+y^2)^3}
\int_{0}^{\infty}dt e^{-t}\left[ a_{2}^{(2)}(z)(\ln t +C)+d(z)\right]
\nonumber \\
&& d(z)=\frac{2}{z} a_{2}^{(1)}(z)-\frac{1}{z^2}
\left( a_{2}(z)-a_{2}(0)\right).
\label{b.6}
\end{eqnarray}
We proceed as above (see Eq.(\ref{a.8})) and expand
$(1+y^2)^3 \simeq 1-3y^2$, then
\begin{eqnarray}
&&  T_{21} = \frac{2}{|\beta|^2} {\rm Im}\int_{0}^{\beta y_0}\frac{dy}{y}
\int_{0}^{\infty}dt e^{-t}\left[ a_{2}^{(2)}
\left(\frac{t}{y}\right) (\ln t +C)+d\left(\frac{t}{y}\right)\right]
\nonumber \\
&&=\frac{1}{|\beta|^2}\frac{1}{i} \int_{\beta^{\ast}y_0}^{\beta y_0}
\frac{dy}{y}\int_{0}^{\infty}dt e^{-t}\left[ a_{2}^{(2)}
\left(\frac{t}{y}\right) (\ln t +C)+d\left(\frac{t}{y}\right)\right]
\nonumber \\
&& \simeq \frac{1}{|\beta|^2}\left[\frac{\pi}{2}d(0)
+\frac{1}{|\beta|^2y_0^2}\left(\frac{3}{2} a_{2}^{(4)}(0)
+d^{(2)}(0) \right) \right]
\nonumber \\
&& = \frac{1}{|\beta|^2}\left[\frac{3\pi}{4}a_{2}^{(2)}(0)
+\frac{25}{12|\beta|^2y_0^2}a_{2}^{(4)}(0)\right].
\label{b.7}
\end{eqnarray}
and
\begin{eqnarray}
&& T_{22} = -\frac{6}{|\beta|^2} {\rm Im}\int_{0}^{y_0}ydy
\int_{0}^{\infty}dt e^{-t}\left[ a_{2}^{(2)}
\left(z\right) (\ln t +C)+d\left(\frac{t}{y}\right)\right]
\nonumber \\
&&= -\frac{6}{|\beta|^2} {\rm Im}\int_{0}^{y_0}ydy
\int_{0}^{\infty}dt e^{-t}\left[ (a_{2}^{(2)}
\left(z\right) -a_{2}^{(2)}(0))(\ln t +C)+d\left(\frac{t}{y}\right)\right].
\label{b.8}
\end{eqnarray}
Integrating by parts over t two times we obtain
\begin{eqnarray}
&& T_{22} = \frac{6}{|\beta|^4} {\rm Re}\int_{0}^{\beta y_0}\frac{dy}{y}
\int_{0}^{\infty}dt e^{-t}\left[ a_{2}^{(4)}
\left(\frac{t}{y}\right) (\ln t +C)+g\left(\frac{t}{y}\right)\right]
\nonumber \\
&& g(z)=\frac{2}{z} a_{2}^{(3)}(z)-\frac{1}{z^2}
\left( a_{2}^{(2)}(z)-a_{2}^{(2)}(0)\right)+d^{(2)}(z).
\label{b.9}
\end{eqnarray}
The term with the function $g(z)$ can be calculated in the same manner
as in Eq.(\ref{a.10}) ($\varphi_{2}^{(4)}(z)\rightarrow g(z)$)
\begin{equation}
T_{22}^{(g)} = \frac{6}{|\beta|^4} {\rm Re}\left[
(\ln (\beta y_0) +C)g(0)- \int_{0}^{\infty}
 \ln z g'(z) dz\right]
\label{b.10}
\end{equation}
Now we have to calculate the integral containing $\ln t +C$
\begin{eqnarray}
&& T_{22}^{(l)} = \frac{6}{|\beta|^4} {\rm Re}\int_{0}^{\beta y_0}\frac{dy}{y}
\int_{0}^{\infty} e^{-t} a_{2}^{(4)}
\left(\frac{t}{y}\right) (\ln t +C)dt
\nonumber \\
&& \simeq \frac{6}{|\beta|^4} \int_{0}^{\infty} \ln y dy
\int_{0}^{\infty} z e^{-zy}(\ln y + \ln z +C)a_{2}^{(5)}(z)dz
\nonumber \\
&& = \frac{6}{|\beta|^4} \int_{0}^{\infty} dy
\int_{0}^{\infty} (\ln y - \ln z) (\ln y + C)e^{-y}a_{2}^{(5)}(z) dz
\nonumber \\
&& = -\frac{6}{|\beta|^4}a_{2}^{(4)}(0)
\int_{0}^{\infty}e^{-y}(\ln^2 y - C^2)dy
= -\frac{\pi^2}{|\beta|^4}a_{2}^{(4)}(0)
\label{b.11}
\end{eqnarray}
Taking into account that $\displaystyle{g(0)=\frac{25}{12}a_{2}^{(4)}(0)}$
and summing $T_1$ Eq.(\ref{b.5}), $T_{21}$ Eq.(\ref{b.7}),
$T_{22}^{(g)}$ Eq.(\ref{b.10}), $T_{22}^{(l)}$ Eq.(\ref{b.11}) we find
for $T$ Eq.(\ref{b.4})
\begin{eqnarray}
\hspace{-10mm}&& T= \frac{3\pi}{4}\frac{\varepsilon}{\varepsilon_e}a_{2}^{(2)}(0)+
\frac{\varepsilon^2}{\varepsilon_e^2}\left[\frac{25}{4}
\left(\ln \frac{\varepsilon_e}{\varepsilon}+2C -\frac{5}{6}
-\frac{4}{25}\pi^2 \right)a_{2}^{(4)}(0)
-6 \int_{0}^{\infty}\ln z g'(z) dz  \right];
\nonumber \\
\hspace{-10mm}&& g(z)=\frac{4}{z}a_{2}^{(3)}(z)-\frac{1}{z^2}
(6a_{2}^{(2)}(z)-a_{2}^{(2)}(0))
+\frac{8}{z^3}a_{2}^{(1)}(z) - \frac{6}{z^4}(a_2(z)-a_2(0)).
\label{b.12}
\end{eqnarray}
Substituting the obtained expression for $T$ Eq.(\ref{b.12}) and $J_{v2}^{(b)}$
Eq.(\ref{b.3}) into $J_{v2}$, adding to $J_{v1}$ and taking into account that
\begin{eqnarray}
&& a_{1}^{(4)}(0)=\frac{1}{30},~b_{1}^{(4)}(0)=\frac{1223}{450},~
a_{2}^{(2)}(0)=-\frac{4}{15},~a_{2}^{(4)}(0)=\frac{16}{21},
\nonumber \\
&& b_{2}(0)=-\frac{1}{18},~b_{2}^{(2)}(0)=-\frac{1}{225},~
b_{2}^{(4)}(0)=\frac{106}{735},
\label{b.13}
\end{eqnarray}
and numerical value of integrals
\begin{equation}
\int_{0}^{\infty}\ln z b_{2}^{(5)}(z) dz \simeq 0.250, \quad
\int_{0}^{\infty}\ln z g'(z) dz \simeq 0.967,
\label{b.14}
\end{equation}
we obtain
\begin{equation}
J_v=\frac{1}{9}+\frac{182\pi}{225}\frac{\varepsilon}{\varepsilon_e}-
\frac{15272}{735}\frac{\varepsilon^2}{\varepsilon_e^2}
\left(\ln \frac{\varepsilon_e}{\varepsilon}-2.577 
\right) +O\left(\frac{\varepsilon^3}{\varepsilon_e^3}\right).
\label{b.15}
\end{equation}

\setcounter{equation}{0}

\section{Appendix}

Here we consider the asymptotic behavior of the total
probability of radiation $W$ in the region
$\varepsilon \ll \varepsilon_e$ to within terms
$\sim \varepsilon/\varepsilon_e$ inclusively.
According with Eqs.(\ref{a.1})-(\ref{a.4}) we present
the total probability in form
\begin{equation}
W=\frac{1}{L_{rad}^0}\left(W_m +\frac{1}{L_1}W_v \right),
\label{c.1}
\end{equation}
where
\begin{eqnarray}
&& W_m=2{\rm Re}\int_{0}^{1}\frac{dx}{x}\int_{0}^{\infty}dt e^{-t}
\left[x^2\varphi_1(z)-2(1-x)\varphi_2(z)\right];
\nonumber \\
&& W_v=-{\rm Re}\int_{0}^{1}\frac{dx}{x}\int_{0}^{\infty}dt e^{-t}
\left[x^2l_1(z, t)+4(1-x)l_2(z, t)\right]
\label{c.2}
\end{eqnarray}
The function entering in Eq.(\ref{c.2}) are defined in
Eqs.(\ref{a.3})-(\ref{a.4}).

The term with $\varphi_1(z)$ in the integrand of $W_m$ to within
mentioned accuracy is
\begin{equation}
W_{m1}=2\int_{0}^{1}x dx\int_{0}^{\infty} e^{-t}dt \varphi_1(0)
\label{c.3}
\end{equation}
In the integral with $\varphi_2(z)$ we divide the integral over
$x$ into two
\begin{equation}
 W_{m2}^{(1)}=-4{\rm Re}\int_{x_0}^{1}\frac{dx}{x}(1-x)
\int_{0}^{\infty}dt e^{-t}\varphi_2(z) \simeq 4(\ln x_0+1)\varphi_2(0),
\label{c.4}
\end{equation}
and
\begin{eqnarray}
&& W_{m2}^{(2)}=-8{\rm Re}\int_{0}^{y_0}\frac{dy}{y(1+y^2)^2}
\int_{0}^{\infty}dt e^{-t}\varphi_2(z)
\simeq -8{\rm Re}\int_{0}^{\beta y_0}\frac{dy}{y}
\int_{0}^{\infty}dt e^{-t}\varphi_2\left( \frac{t}{y}\right)
\nonumber \\
&& +16{\rm Re}\int_{0}^{y_0} y dy
\int_{0}^{\infty}dt e^{-t}\varphi_2\left( \frac{t}{\beta y}\right)
=  W_{m21}^{(2)} +  W_{m22}^{(2)}
\label{c.5}
\end{eqnarray}
Integration of $W_{m21}^{(2)}$ coincides with calculation in
Eq.(\ref{a.10}) with substitution
$\displaystyle{\varphi_2^{(4)}\left( \frac{t}{y}\right) \rightarrow
\varphi_2\left( \frac{t}{y}\right)}$. We have
\begin{eqnarray}
&& W_{m21}^{(2)}= -8{\rm Re}\int_{0}^{\beta y_0}\frac{dy}{y}
\int_{0}^{\infty}dt e^{-t}\varphi_2\left( \frac{t}{y}\right)
\nonumber \\
&& \simeq -4\left[\left(\ln \frac{\varepsilon_e}{\varepsilon}+2C
+\ln y_0^2\right)\varphi_2(0) -2\int_{0}^{\infty}
\ln z \varphi_2^{(1)}(z)dz\right]
\label{c.6}
\end{eqnarray}
Integrating by parts twice over $t$ in integral $W_{m22}^{(2)}$
Eq.(\ref{c.5}) we find (see Eq.(\ref{a.9}))
\begin{equation}
 W_{m22}^{(2)}=16\frac{\varepsilon}{\varepsilon_e}{\rm Im}
\int_{0}^{\beta y_0}\frac{dy}{y}
\int_{0}^{\infty}dt e^{-t}\varphi_2^{(2)}\left( \frac{t}{y}\right)
\simeq 4\pi\frac{\varepsilon}{\varepsilon_e}\varphi_2^{(2)}(0).
\label{c.7}
\end{equation}
Summing $W_{m2}^{(1)}$ Eq.(\ref{c.4}), $W_{m21}^{(2)}$
Eq.(\ref{c.6}) and $W_{m22}^{(2)}$ Eq.(\ref{c.7}) and taking into
account Eq.(\ref{a.12})) we have
\begin{eqnarray}
&& W_{m}= -4\varphi_2(0)\left[\ln \frac{\varepsilon_e}{\varepsilon}
+2C-1 -\frac{\varphi_1(0)}{4\varphi_2(0)}-
\frac{2}{\varphi_2(0)}\int_{0}^{\infty}
\ln z \varphi_2^{(1)}(z)dz\right]+
\nonumber \\
&& 4\pi \frac{\varepsilon}{\varepsilon_e} \varphi_2^{(2)}(0)
 =\frac{4}{3}\left[\ln \frac{\varepsilon_e}{\varepsilon}
+2C- \frac{5}{8}+12\int_{0}^{\infty}\ln z\left(\frac{1}{z^3}-
\frac{\cosh z}{\sinh^3z}\right)dz\right] +\frac{8\pi}{15}
\frac{\varepsilon}{\varepsilon_e}
\nonumber \\
&&= \frac{4}{3}\left(\ln \frac{\varepsilon_e}{\varepsilon}
+1.959\right)+\frac{8\pi}{15}
\frac{\varepsilon}{\varepsilon_e}
\label{c.8}
\end{eqnarray}
The first term in Eq.(\ref{c.8})) is similar to Eq.(2.21) in
\cite{L3}.

We will consider now the probability $W_v$ in Eq.(\ref{c.2})).
The term with $l_1(z,t)$ gives no contribution into two first terms
of expansion over $\displaystyle{\frac{\varepsilon}{\varepsilon_e}}$.
\begin{equation}
W_{v2}=-4{\rm Re}\int_{0}^{1}(1-x)\frac{dx}{x}\int_{0}^{\infty}dt e^{-t}
\left[a_{2}(z)(\ln t +C) + b_{2}(z)\right]=W_{v2}^{(a)}+W_{v2}^{(b)}.
\label{c.9}
\end{equation}
The calculation of the term $W_{v2}^{(b)}$ (with $b_2(z)$) is analogous
to the calculation above (with substitution
$\varphi_2(z) \rightarrow b_2(z)$)
\begin{equation}
W_{v2}^{(b)}=-4 b_2(0)\left(\ln \frac{\varepsilon_e}{\varepsilon}+2C
-1\right) +8\int_{0}^{\infty}
\ln z b_2^{(1)}(z)dz + 4\pi\frac{\varepsilon}{\varepsilon_e}b_2^{(2)}(0).
\label{c.10}
\end{equation}

For calculation of term $W_{v2}^{(a)}$ (with $a_2(z)$) it is
convenient to use the variable $y$
\begin{eqnarray}
\hspace{-10mm}&& W_{v2}^{(a)}=W_{v21}^{(a)}+W_{v22}^{(a)};
\nonumber \\
\hspace{-10mm}&& W_{v21}^{(a)}= -8\int_{0}^{\infty}\frac{dy}{y}
\int_{0}^{\infty}dt e^{-t}\left(a_2\left( \frac{t}{y}\right)-
a_2\left( \frac{1}{y}\right)\right)(\ln t+C),
\nonumber \\
\hspace{-10mm}&& W_{v22}^{(a)}=8{\rm Re}\int_{0}^{\infty}\frac{dy}{y}
\left(1-\frac{1}{(1+y^2)^2}\right)
\int_{0}^{\infty}dt e^{-t}(a_2(z)-a_2(0))(\ln t+C).
\label{c.11}
\end{eqnarray}
The integral over $y$ in $W_{v21}^{(a)}$ is the integral Frullani
\begin{equation}
\int_{0}^{\infty}\left(a_2\left(\frac{t}{y} \right)
-a_2\left(\frac{1}{y} \right) \right) \frac{dy}{y}=-a_2(0)\ln t.
\label{c.12}
\end{equation}
Using the last result in Eq.(\ref{b.11})) we have
\begin{equation}
W_{v21}^{(a)}=\frac{4}{3}\pi^2a_2(0).
\label{c.13}
\end{equation}
In the expression for $W_{v22}^{(a)}$ we divide the integration interval
into two parts:
\begin{eqnarray}
&& W_{v22}^{(a1)}=8{\rm Re}\int_{y_0}^{\infty}\frac{dy}{y}
\left(1-\frac{1}{(1+y^2)^2}\right)
\int_{0}^{\infty} e^{-t}\left( a_2^{(2)}(0)
\frac{t^2}{2!\beta^2y^2}+...\right) (\ln t+C)dt=0,
\nonumber \\
&& W_{v22}^{(a2)} \simeq 16{\rm Re}\int_{0}^{y_0} ydy
\int_{0}^{\infty} e^{-t}(a_2(z)-a_2(0))(\ln t+C)dt
\nonumber \\
&& = 16{\rm Im}\int_{0}^{\beta y_0}\frac{dy}{y}
\int_{0}^{\infty} e^{-t}\left[a_2^{(2)}\left(\frac{t}{y}
\right)(\ln t+C) + d\left(\frac{t}{y}\right) \right]dt
\label{c.14}
\end{eqnarray}
The function $d(z)$ is defined in Eq.(\ref{b.6}), the further
calculation can be done as in Eq.(\ref{b.7})
\begin{equation}
W_{v22}^{(a2)} \simeq 4\pi \frac{\varepsilon}{\varepsilon_e}d(0)
=6\pi \frac{\varepsilon}{\varepsilon_e}a_2^{(2)}(0).
\label{c.15}
\end{equation}
Summing $W_{v2}^{(b)}$ Eq.(\ref{c.10}), $W_{v21}^{(a)}$ Eq.(\ref{c.13})
and  $W_{v21}^{(a)}$ Eq.(\ref{c.15}) we obtain the following
expression for $W_v$ Eq.(\ref{c.1})
\begin{equation}
W_v=-4b_2(0)\left(\ln \frac{\varepsilon_e}{\varepsilon} +2C-1 \right)
+8 \int_{0}^{\infty}\ln z b_2^{(1)}(z)dz +\frac{4\pi^2}{3}
+2\pi\frac{\varepsilon}{\varepsilon_e}(3a_2^{(2)}(0) + 2b_2^{(2)}(0)).
\label{c.16}
\end{equation}
Taking into account that
\[
\int_{0}^{\infty}\ln z b_2^{(1)}(z)dz = 0.0431
\]
and using Eq.(\ref{b.13}) we obtain ($a_2(0)=1/3$)
\begin{equation}
W_v= \frac{2}{9}\ln \frac{\varepsilon_e}{\varepsilon}
+4.7655 -\frac{364\pi}{225}
\frac{\varepsilon}{\varepsilon_e} +
O\left(\frac{\varepsilon^2}{\varepsilon_e^2} \right).
\label{c.17}
\end{equation}

\setcounter{equation}{0}

\section{Appendix}

Here we discuss a structure of corrections to the total intensity
of radiation at $\varepsilon \ll \varepsilon_e$ which is not
included in the expansion over $1/L_c$ contained in
Eqs.(\ref{3.1})- (\ref{3.3}). We will show that these corrections,
which are contained terms $\propto 1/L_1^2$ and higher degrees of
$1/L_1$, are beginning from terms of the order
$\varepsilon/\varepsilon_e$. We will use the general expression
Eq.(2.12) of \cite{L1} before decomposition over $1/L$:
\begin{equation}
\frac{dI}{dx}=\frac{2\alpha m^2 x}{1-x}{\rm Im}\left<  0|r_1
\left(G^{-1}-G_0^{-1}\right) +r_2{\bf p}
\left(G^{-1}-G_0^{-1}\right){\bf p} |0\right>,
\label{d.1}
\end{equation}
where
\begin{equation}
G={\bf p}^2+1-iV(\mbox{\boldmath$\varrho$}),~ G_0={\bf p}^2+1,~
V(\mbox{\boldmath$\varrho$})=\frac{\varepsilon}{4\varepsilon_e}
\frac{1-x}{x}\mbox{\boldmath$\varrho$}^2\left( L_1-\ln
\frac{\mbox{\boldmath$\varrho$}^2}{4}-2C\right),
\label{d.2}
\end{equation}
here the quantities $r_1,r_2$ are defined in Eq.(\ref{3.2}) and
$\varepsilon_e$ and $L_1$ are defined in Eq.(\ref{3.3}). In the
region $1 \geq x \gg \varepsilon/\varepsilon_e$ we expand the
combination entering in Eq.(\ref{d.1}) over degrees of $V$
\begin{equation}
G^{-1}-G_0^{-1}=G_0^{-1}iVG_0^{-1}+G_0^{-1}iVG_0^{-1}iVG_0^{-1}+
G_0^{-1}iVG_0^{-1}iVG_0^{-1}iVG_0^{-1}+\dots
\label{d.3}
\end{equation}
Substituting Eq.(\ref{d.3}) into Eq.(\ref{d.1}) one can verify
that the terms of the order $ 1/L_1^2$ and higher degrees of
$1/L_1$ are appeared starting from the third term of decomposition
Eq.(\ref{d.3}). Their relative contribution is of the order
$\varepsilon^2/\varepsilon_e^2$. Taking into account that
\begin{equation}
G_0^{-1}=i\int_{0}^{\infty}e^{-it} e^{-i{\bf p}^2t}dt,~
<\mbox{\boldmath$\varrho$}_1|e^{-i{\bf
p}^2t}|\mbox{\boldmath$\varrho$}_2>=\frac{1}{4\pi
it}e^{\frac{i}{4t}\left(\mbox{\boldmath$\varrho$}_1-
\mbox{\boldmath$\varrho$}_2\right)^2}, \label{d.4}
\end{equation}
we obtain
\begin{equation}
<\mbox{\boldmath$\varrho$}_1|G_0^{-1}
|\mbox{\boldmath$\varrho$}_2> =
\frac{1}{4\pi}\int_{0}^{\infty}\exp\left[-it
+\frac{i}{4t}(\mbox{\boldmath$\varrho$}_1
-\mbox{\boldmath$\varrho$}_2)^2\right]\frac{dt}{t}=
\frac{1}{2\pi}K_0(|\mbox{\boldmath$\varrho$}_1-\mbox{\boldmath$\varrho$}_2|),
\label{d.5}
\end{equation}
where $K_0(z)$ is the modified Bessel function (Mac-Donald's
function). Substituting this result into Eq.(\ref{d.3}) and then
into Eq.(\ref{d.1}) we get
\begin{eqnarray}
&& \frac{dI}{dx}=\frac{\alpha
m^2x}{\pi(1-x)}\int_{0}^{\infty}\left[
r_1K_0^2(\varrho)+r_2K_1^2(\varrho) \right]V(\varrho)\varrho
d\varrho
\nonumber \\
&&=\frac{\alpha m^2}{4\pi}\frac{\varepsilon}{\varepsilon_e}
\left[\frac{r_1+2r_2}{3}L_1+\frac{r_2-r_1}{9}\right],\quad
I^{(0)}=\frac{\alpha
m^2}{4\pi}\frac{\varepsilon}{\varepsilon_e}\left(L_1+
\frac{1}{9}\right).
\label{d.6}
\end{eqnarray}
Obtained here result for $I^{(0)}$ agrees with Bethe-Maximon
formula.

In the region $x \sim \varepsilon/\varepsilon_e$ the potential $V
\sim 1$ and decomposition Eq.(\ref{d.3}) is inapplicable. However
in this region
\begin{equation}
dI \sim \alpha m^2\frac{xdx}{1-x} \sim  \alpha m^2
\left(\frac{\varepsilon}{\varepsilon_e}\right)^2,
\label{d.7}
\end{equation}
and the relative contribution of this region $\sim
\varepsilon/\varepsilon_e$.

\newpage

{\bf Figure captions}

\vspace{8mm}
\begin{itemize}

\item {\bf Fig.1} 
The energy loss spectrum
$\displaystyle{\frac{d\varepsilon}{d\omega}}$
in units $\displaystyle{\frac{1}{L_{rad}^0}}$
in iridium target with thickness $l$=0.128~mm 
($ l/L_{rad}^{BM}=4.36 \%$) for the initial electrons 
energy is $\varepsilon=286.6~GeV$.
The Coulomb corrections are included.
\begin{itemize}
\item Curve 1 is the Bethe-Maximon spectrum,
\item curve 2 is the contribution of the main term describing LPM effect;
\item curve 3 is the correction term;
\item curve 4 is the sum of two previous contributions;
\item curve T is the final theory prediction with regard for
the reduction factor (the multiphoton effects).
\end{itemize}
Experimental data from \cite{HU}, \cite{HU1}.

\item {\bf Fig.2} The same as in Fig.1 but for 
the initial electrons energy $\varepsilon=206.7~GeV$.

\item {\bf Fig.3} The same as in Fig.1 but for tantalum target
with thickness ($ l/L_{rad}^{BM}=4.45 \%$).

\item {\bf Fig.4} The reduction factor for iridium target 
with thickness $l=0.128~mm$ (4.36 \% $L_{rad}$) versus 
$x=\omega/\varepsilon$. The curves 1, 2, 3 are for energies
149~GeV, 207~GeV and 287~GeV correspondingly.

\item {\bf Fig.5} The relative energy losses of electron
per unit time including the contribution of the correction term 
$(I(\varepsilon)/\varepsilon) L_{rad}^{BM}=L_{rad}^{BM}/L_{rad}$
in iridium (curve 1) and in lead (curve 2) vs the initial energy
of electron.

\end{itemize}

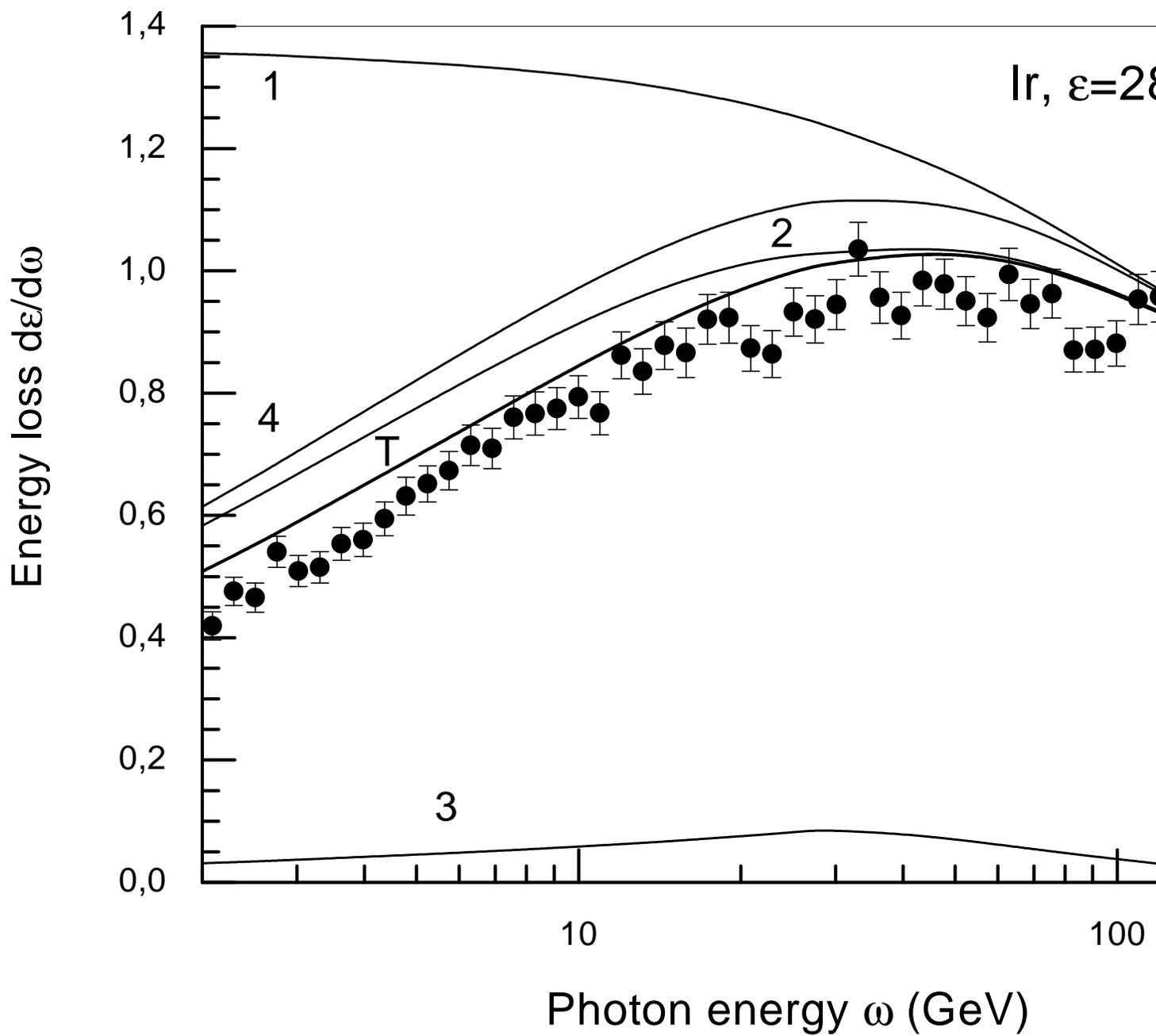
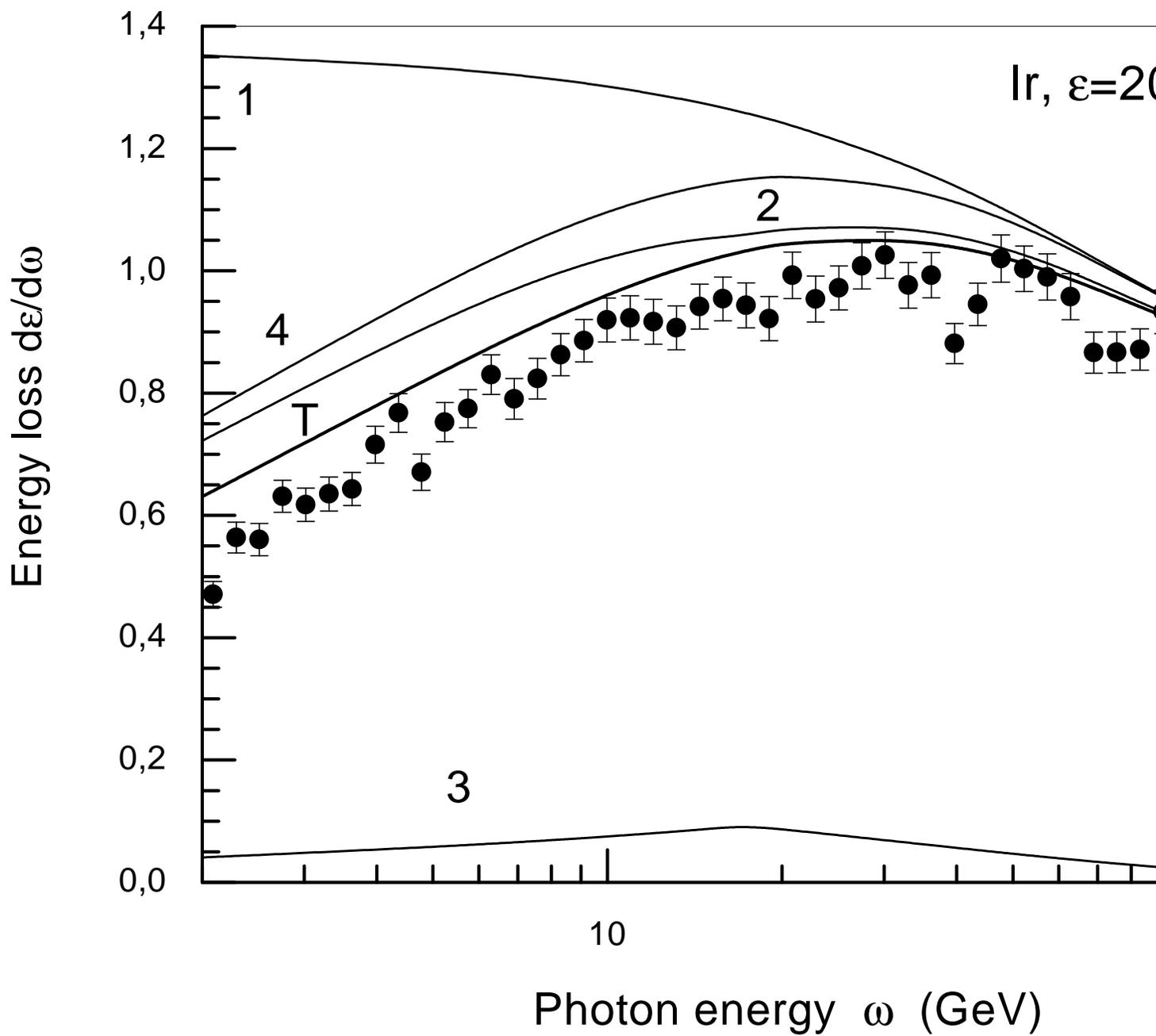
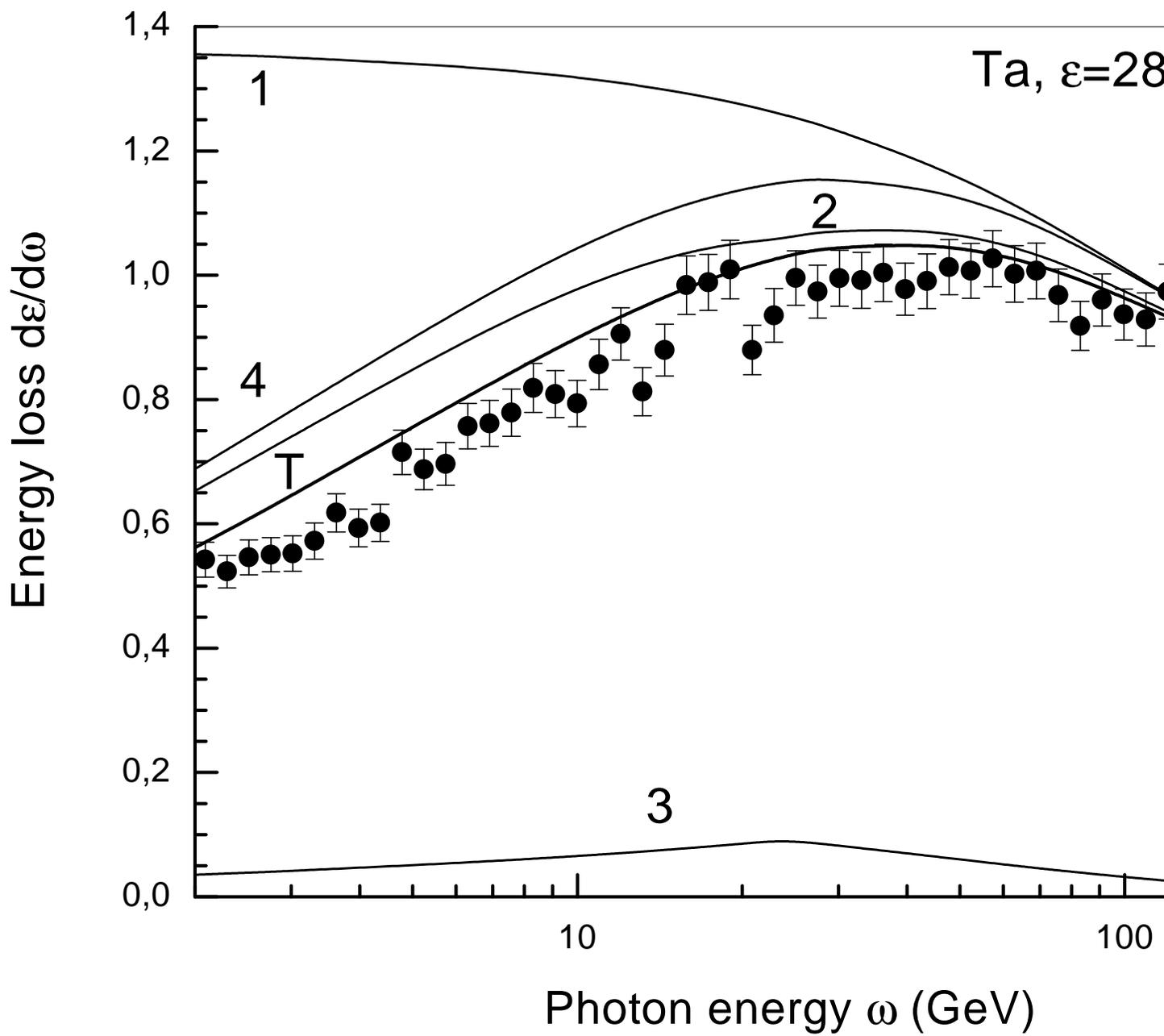
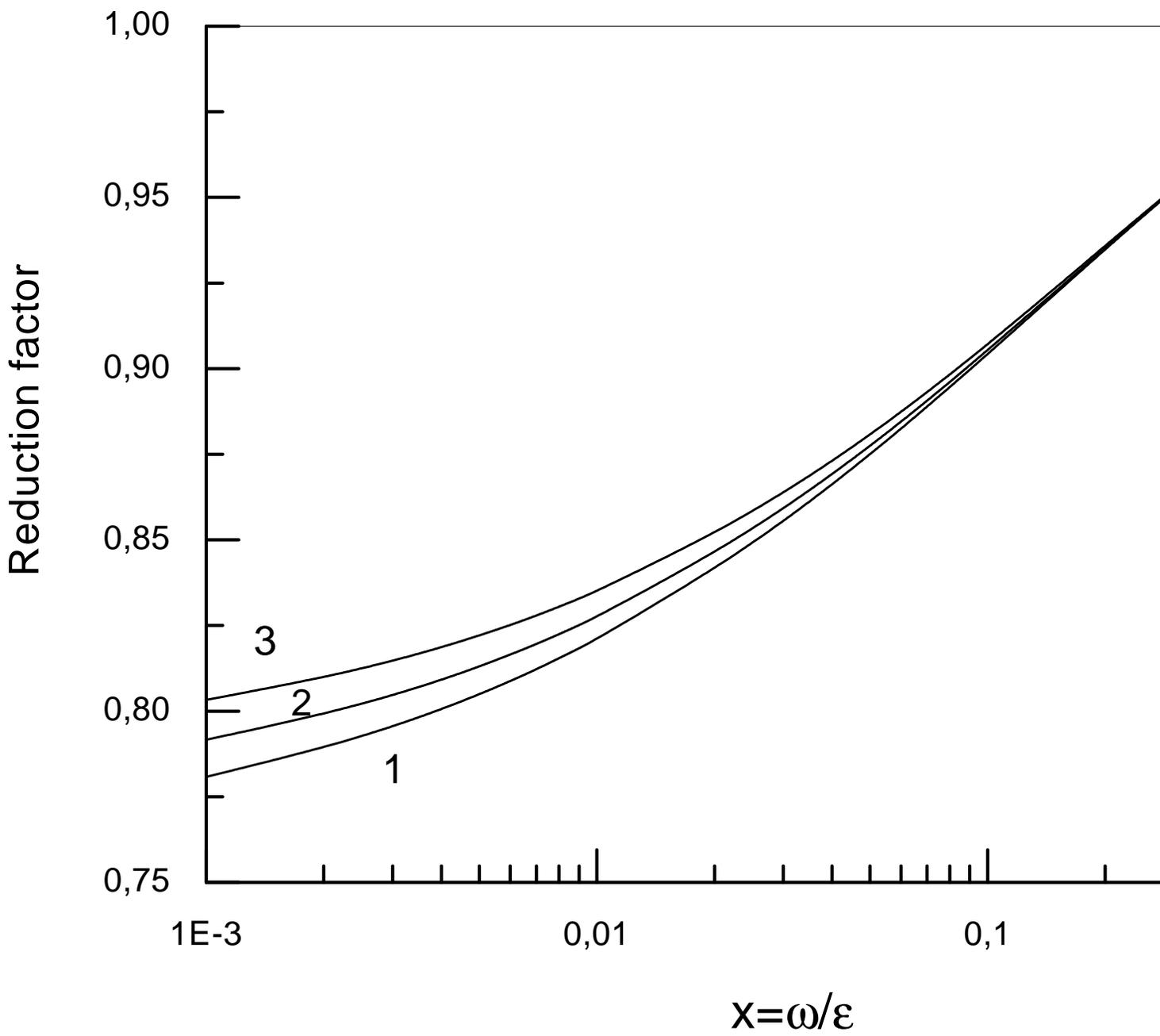
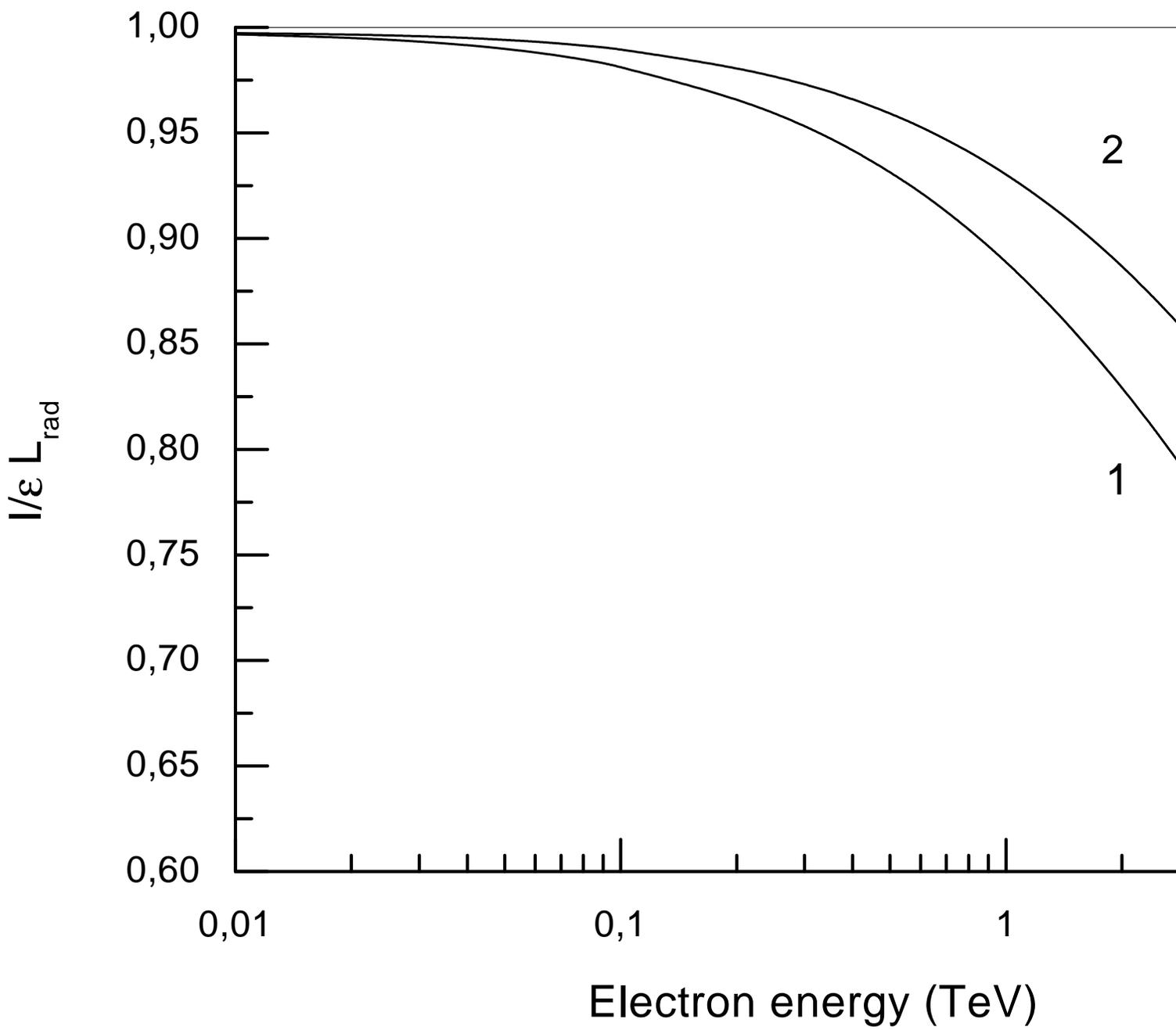
\end{document}